\documentclass{aa}  

\usepackage{graphicx}
\usepackage{txfonts}
\usepackage{hyperref}
\usepackage{verbatim}
\usepackage{mypckg}
\usepackage{xcolor}
\begin{document}

   \title{Detailed seismic study of Gemma (KIC11026764) using \egg}

   \subtitle{Unveiling the probing potential of mixed modes for subgiant stars}

   \author{M. Farnir\inst{1}
          \and
          M.-A. Dupret\inst{1}
          \and
          G. Buldgen\inst{1}
          }

   \institute{STAR Institute, Université de Liège, Liège, Belgium \email{martin.farnir@uliege.be}
             }

   \date{Received Month DD, YYYY; accepted Month DD, YYYY}

 
  \abstract
   {When leaving the main sequence (MS) for the red-giant branch (RGB), subgiant stars undergo fast structural changes. Consequently, their observed oscillation spectra mirror these changes, constituting key tracers of stellar structure and evolution. However, the complexity of their spectra makes their modelling an arduous task, which few authors have undertaken. Gemma (KIC11026764) is a young subgiant with $45$ precise oscillation modes observed with Kepler, making it the ideal benchmark for seismic modelling.}
   {This study is aimed at modelling the subgiant Gemma, taking advantage of most of the precise seismic information available. This approach enables us to pave the way for the seismic modelling of evolved solar-like stars and provide the relevant insights into their structural evolution.}
   {Using our Levenberg-Marquardt stellar modelling tool, we built a family of models representative of Gemma's measured seismic indicators obtained via our seismic tool, EGGMiMoSA. We studied the structural information these indicators hold by carefully varying stellar parameters. We also complemented the characterisation with information held by \who indicators and non-seismic data.}
   {From the extensive set of models we built and using most of the seismic information at hand, including two $\ell=1$ and one $\ell=2$ mixed modes, we were able to probe the chemical transition at the hydrogen-burning shell. Indeed, we have demonstrated that among our models, only the ones with the sharpest chemical gradient are able to reproduce all the seismic information considered. One possibility to account for such a gradient is the inclusion of a significant amount of overshooting, namely $\alpha_{\textrm{ov}}=0.17$, which is unexpected for low-mass stars such as Gemma (expected mass of about $1.15~M_{\odot}$).}
   {}

   \keywords{stars:subgiants stars:individual:KIC11026764
               }

   \maketitle
%

\section{Introduction}
Because of the large distances involved, the only information that can be gathered about stars is carried by their light. However, due to the opacity of the stellar interior, the only information that can be retrieved is related to the superficial layers. Nevertheless, constraining the deeper regions of a star is necessary to improve our models and we can do so through asteroseismology, the study of stellar oscillations and their relation with the stellar structure. A very efficient asteroseismic framework is the asymptotic theory \citep{1980ApJS...43..469T,1989nos..book.....U} which identifies two kinds of oscillation (according to their restoring forces) that both display regularity in their spectra: the pressure modes (henceforth p-modes), equally spaced in frequency with the pressure gradient acting as the restoring force, and the gravity modes (g-modes), equally spaced in period with the buoyancy being the restoring force. This regularity (either in frequency or period) is expressed through two numbers: $\ell$, the spherical degree, and $n$, the radial order (with either a subscript p or g for p or g-modes)\footnote{If the rotation were to be considered, a third number would be included: the azimuthal order $m$.}. These two kinds of modes are associated with characteristic frequencies, which allow to determine the regions they are able to probe. Here, we have:\\
\begin{equation}
    S_\ell^2 = \frac{\ell \left(\ell+1\right) c^2}{r^2} ,
\end{equation}
the (squared) Lamb frequency associated to p-modes of spherical degree $\ell$: with $r$ as the distance from the stellar center, and $c$ the local sound speed, and
\begin{equation}
N^2 = \frac{g\rho_T}{H_p}\left(\nabla_{\textrm{ad}}-\nabla + \frac{\rho_\mu}{\rho_T}\nabla_{\mu} \right), \label{Eq:Brunt}
\end{equation}
is the (squared) Brunt-Väisälä frequency associated with the g-modes: with $g$ the local gravity, $H_p$ the pressure scale height, $\rho_T=-\left.\partial\ln \rho/\partial \ln T\right\vert_{P,\mu}$, $\rho_\mu=\left.\partial\ln \rho/\partial\ln\mu\right\vert_{P,T}$, $\nabla=\partial\ln T/\partial\ln P$, $\nabla_{\textrm{ad}}=\left.\partial\ln T/\partial\ln P\right\vert_{\textrm{ad}}$, $\nabla_\mu=\partial\ln \mu/\partial\ln P$, $\rho$ the local density, $T$ the temperature, $P$ the pressure, and $\mu$ the mean molecular weight.
In low-mass main sequence stars, the two types of modes propagate in two separate cavities. The g-modes probe the deeper regions of stars, while p-modes propagate in the outer regions. Therefore, they hold complementary information. Unfortunately, g-modes have yet to be detected for MS solar-like stars, leaving the deepest layers outside of our reach.

As these stars evolve to the subgiant and then red-giant phases, their cores contract and their envelopes expand. This creates a strong density contrast. Along with it, the frequencies of the p-modes (in the outer regions) decrease and those of the g-modes (in the innermost regions) increase. As both frequencies get close to one another, modes exchange nature and we observe avoided crossings \citep[See for example ][\href{https://www.aanda.org/articles/aa/pdf/2011/11/aa17232-11.pdf\#figure.1}{Fig.1}]{2011A&A...535A..91D}. Additionally, due to the important density contrast in these stars, the p- and g-modes are coupled, creating peculiar modes referred to as mixed-modes. These propagate throughout their entire structure, alternating a g-dominated nature in the core and a p-dominated nature at the surface.

The existence of such modes is a unique opportunity to probe missing physical processes down to their deepest regions (this is not the case for their main-sequence cousins, such as our Sun for which only superficial p-modes are detected). It was previously demonstrated, for example, that subgiant and then red-giant stars present cores that rotate much slower than what would be expected from standard theories, up to two orders of magnitude \citep{2012A&A...548A..10M,2014A&A...564A..27D,2020A&A...641A.117D,2018A&A...616A..24G,2024A&A...689A.307B}. To compensate for this discrepancy, several angular momentum redistribution mechanisms have been proposed, including transport via internal gravity waves \citep{2017A&A...605A..31P} or magnetic fields \citep[and references therein]{2019MNRAS.485.3661F,2019A&A...631L...6E,2022A&A...661A.119G,2024A&A...681A..75P}. However, in most cases, angular momentum transport studies are decoupled from stellar structure characterisation and a dedicated modelling of a broad sample of evolved solar-like stars remains necessary to constrain these processes and draw meaningful statistical inferences. 

Another important aspect of the stellar structure is the mixing beyond the boundaries of convective regions. The amount and exact nature of this extra mixing, usually depicted by a single ad-hoc parameter, remains an open question which several authors have attempted to answer. For example, \citet{2016A&A...589A..93D} used asteroseismology to constrain the overshooting in MS low-mass stars and observed a correlation between the overshooting parameter and mass. For later stages of evolution, \citet{2013A&A...549A..74M,2015ApJ...805..127C,2015MNRAS.453.2290B,2017MNRAS.469.4718B} were able to show that the regularity in period of the modes present in core helium-burning stars allows us to put constraints on the core mixing. \citet{2021A&A...650A.115D} showed that the overshooting parameter is a function of both the mass and (to a lesser extent) the metallicity of AGB stars. Finally, for the case of a specific subgiant star, KIC10273246, \citet{2021A&A...647A.187N} inferred a mass of $1.31\pm 0.03 M_{\odot}$ (significantly more massive than the expected $1.18 \pm 0.08M_{\odot}$. They also found that two g-dominated mixed modes are sufficient to constrain the overshooting parameter and a value of $\alpha_{\textrm{ov}}~=~0.15$ allowed them to reproduce at best observed frequencies. Several authors have taken preliminary steps towards the modelling of subgiant stars mainly by relying on large grids of models and adjusting individual oscillation frequencies \citep[e.g.][]{2014MNRAS.444.3622J,2018A&A...610A..80H,2020MNRAS.495..621J,2020MNRAS.495.3431L,2024ApJ...965..171L}. However, detailed studies are necessary using carefully selected seismic indicators that contain identifiable structural information. 
\citet{2021A&A...653A.126F} have developed a technique that relies on the asymptotic description of the mixed-mode pattern \citep{1979PASJ...31...87S,2015A&A...584A..50M} to adjust the observed and modelled spectra and to define related seismic indicators. They theoretically identified $\Delta\Pi_1$ and $\Delta\nu_0$ as promising indicators to constrain both the mass and age of given target. 

In the present paper, we provide a detailed characterisation of the subgiant Gemma (KIC11026764) using the measured $\Delta\nu_0$ and $\Delta\Pi_1$ pair obtained from the frequencies provided in \citet{2020MNRAS.495.2363L} for several physical choices. Our favoured model shows a significant composition gradient located at the hydrogen burning shell (henceforth H-shell), which we reproduce using significant overshooting. Additionally, we also characterised the agreement of our models with other seismic \citep[helium glitch amplitude, as computed with \who ][]{2019A&A...622A..98F} and non-seismic data (effective temperature, surface gravity, and metallicity). This constitutes a pilot study that is meant to demonstrate that our method yields robust and plausible results. Additional subgiant stars will be considered in a follow-up paper to infer trends between fundamental parameters.

\section{Observed data}
\begin{figure}
    \centering
    \includegraphics[width=.95\linewidth]{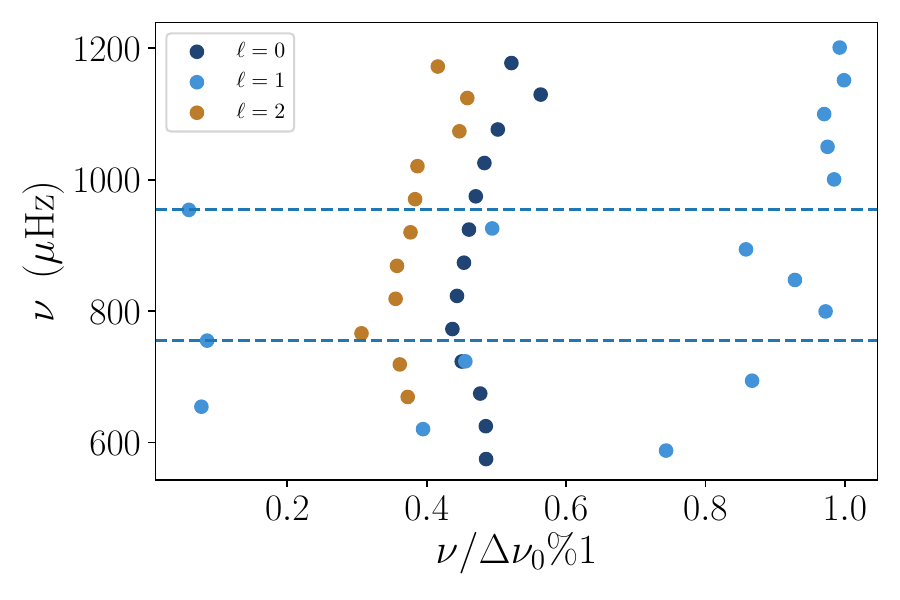}
        \caption{KIC11026764's \emph{échelle} diagram from the frequencies measured by \citet{2020MNRAS.495.2363L}. The large frequency separation of radial modes determined by WhoSGlAd is used. Dark blue symbols correspond to radial modes, the light blue ones to dipolar modes and, the brown ones to quadrupolar modes. The horizontal dashed lines highlight the position of the detected g-dominated dipolar mixed modes.}
    \label{Fig:Ech}
\end{figure}

Gemma (KIC11026764) is a subgiant that is very similar to our Sun -- $\left[Fe/H\right]~=~0.04 \pm 0.15$, $\log g~=~3.89 \pm 0.06$, and $T_{\textrm{eff}}~=~5636 \pm 80$ \citep{2012A&A...543A..54A} --. It has been observed by the Kepler spacecraft \citep{2010AAS...21510101B} for $985$ days, resulting in $45$ detected oscillation modes spanning spherical degrees, $\ell$, from $0$ to $3$ \citep{2020MNRAS.495.2363L}\footnote{\citet{2012A&A...543A..54A} also provided frequencies for Gemma but the range of detected modes is reduced in comparison and only one dipolar mixed mode was detected, impairing the determination of our seismic indicators.}. Among these, at least two g-dominated dipolar mixed modes can be clearly identified from inspection of the \emph{échelle} diagram shown in Fig.~\ref{Fig:Ech}. Indeed, we observe that several $\ell=1$ modes (in light blue) have separated themselves from the vertical ridge expected from a purely pressure mode spectrum. The identified g-dominated modes have been highlighted with horizontal dashed lines. We also identified one quadrupolar g-dominated mode at $766.24~\left(\mu\textrm{Hz}\right)$. This makes Gemma an ideal benchmark for a detailed seismic characterisation. Therefore, we computed Gemma's seismic indicators using \egg and retrieved the signature of the helium glitch on radial modes using \who. These indicators are: the radial modes large frequency separation $\Delta\nu_0$, the dipolar modes period spacing $\Delta\Pi_1$, the pressure offset $\epsilon_{p,1}$, the gravity offset $\epsilon_{g,1}$, the coupling constant $q_1$, the helium glitch amplitude $\mathcal{A}_{\textrm{He}}$, and the small separation (Eq.~\ref{Eq:d02}) between the quadrupolar g-dominated mode and the closest radial mode $d_{02}$. The relevant constraints are presented in Table~\ref{Tab:obsPar}. We draw the attention to the fact that the uncertainties on the measured seismic indicators are unrealistically small as they only are the result of error propagation through the \egg algorithm. They reflect only the observed uncertainties on the individual modes. We also add that we do not provide a value for the luminosity of Gemma as there is no (to our knowledge) literature determination available. In addition, due to possible contamination, it is not possible to compute a reliable estimate of Gemma's luminosity using the latest Gaia data release \citep[DR3][]{refId0}.

\setlength{\tabcolsep}{4pt} 

\begin{table}[h]
    \centering
    \caption{Summary of the relevant observables for Gemma. When no reference is provided, the quantities were derived with \who or \egg using \citet{2020MNRAS.495.2363L}'s set of modes.}
    \begin{tabular}{c|c|c}
    Quantity & Value & Source \\
    \hline
    \hline
      $\log g$ & $3.89 \pm 0.06$ & 1 \\
      $T_{\textrm{eff}}$ (K) & $5636 \pm 80$ & 1 \\
      $\left[Fe/H\right]$ (dex) & $0.04 \pm 0.15$ & 1 \\
      $\Delta\nu_0$ $\left(\mu Hz\right)$ & $50.031 \pm 0.001$ & \who \\
      $\mathcal{A}_{\textrm{He}}$ $\left(\mu Hz\right)$ & $0.0048 \pm 0.0004$ & \who \\
      $\Delta\Pi_1$ (s) & $269.8200 \pm 0.0001$ & \egg \\
      $\epsilon_{p,1}$ & $0.9388 \pm 0.0004$ & \egg \\
      $\epsilon_{g,1}$ & $0.995 \pm 0.002$ & \egg \\
      $q_1$ & $0.1628 \pm 0.0008$ & \egg \\
      $d_{02,\textrm{g}}~\left(\mu \textrm{Hz}\right)$ & $6.53 \pm 0.08$ & {Sect.~\ref{Sec:Quad}} \\
    \end{tabular}
    \tablefoot{The uncertainties returned by the WhoSGlAd and EGGMiMoSA come from error propagation over the observed frequencies. As such, they tend to be overoptimistic.}
    \tablebib{(1)~\citet{2012A&A...543A..54A}}
    \label{Tab:obsPar}
\end{table}

\setlength{\tabcolsep}{6pt} 

\section{Methodology}
\subsection{\egg}
Before detailing the modelling procedure, let us recall the basic principle behind the \egg method \citep{2021A&A...653A.126F}. To extract relevant seismic indicators, the approach adjusts the asymptotic formulation for mixed-mode spectra (expressed here for dipolar modes) derived by \citet{1979PASJ...31...87S} and later adapted by \citet{2015A&A...584A..50M}:
\begin{equation}
    \tan \theta_p = q_1 \tan \theta_g,
    \label{Eq:AsyCou}
\end{equation}
with 
\begin{equation}
    \theta_p = \pi \left[\frac{\nu}{\Delta\nu_1}-\frac{1}{2}-\epsilon_{p,1}\right]
\end{equation}
as the pressure phase and
\begin{equation}
    \theta_g = \pi \left[\frac{1}{\nu \Delta \Pi_1}-\epsilon_{g,1}\right]\label{Eq:tg}
\end{equation}
as the gravity phase, respectively describing the contributions of the pressure and gravity parts of a given mixed mode, respectively. 
Equation~\ref{Eq:AsyCou} is an implicit function of the frequency, $\nu$, and contains five parameters to be adjusted: the large frequency separation $\Delta\nu_1$, the period spacing $\Delta\Pi_1$, the pressure offset $\epsilon_{p,1}$, the gravity offset $\epsilon_{g,1}$, and the coupling constant $q_1$. Relying on educated guesses, the \egg method provides a Levenberg-Marquardt adjustment of these parameters to find the theoretical best-fit spectrum, representative of the observed frequencies. These parameters constitute robust indicators to probe the stellar structure as demonstrated in \cite{2021A&A...653A.126F}.

\subsection{\who}
As we directly use the large frequency separation of radial modes obtained via the \who method \citep{2019A&A...622A..98F}, $\Delta\nu_0$, as a constraint in our modelling procedure and the helium glitch amplitude, $\mathcal{A}_{\textrm{He}}$, to discriminate between optimal models, we provide a short reminder of the method. More detailed information may be found in \citet{2019A&A...622A..98F,2023MNRAS.tmp..738F}.

The strength of the \who method comes from the use of an orthonormal basis of functions to represent the observed frequencies. These functions are separated into a smooth -- slowly varying trend as a function of frequency \citep[as expected from the asymptotic theory: e.g.][]{1980ApJS...43..469T} -- and a glitch contribution, which are mathematically independent of one another by construction. In this formalism, the general representation of a fitted frequency of radial order $n$ and spherical degree $\ell$ is the following
\begin{equation}
    \nu_{n,\ell,\textrm{fit}} = \sum\limits_k a_{k} q_k\left(n\right),
\end{equation}
with $a_{k}$ the projected reference frequency over the basis function of index $k$, $q_k$, evaluated at $n$. These orthonormal functions are obtained by applying the Gram-Schmidt algorithm to the polynomials of increasing degrees $n^k$ for the smooth part and a parametrised oscillating component for the glitch. The projection is done for each spherical degree according to the scalar product defined in \cite{2019A&A...622A..98F}. The specificity of this definition of the scalar product is that it accounts for the observational uncertainties on the oscillation frequencies.

One benefit of the orthonormalisation is that the $a_k$ coefficients are completely independent of one another and of unit uncertainty. Therefore, they can be combined judiciously in order to construct indicators that are as little correlated as possible. The large frequency separation of radial modes and helium glitch amplitude are two such indicators that are completely uncorrelated from one another. We briefly recall their mathematical expressions \citep[further information may be found in ][]{2019A&A...622A..98F,2023MNRAS.tmp..738F}. We first define the large frequency separation associated to modes of spherical degree $\ell$ as:
\begin{equation}\label{Eq:Who-DnulWho}
\Delta\nu_\ell = a_{\ell,1} R^{-1}_{\ell,1,1},
\end{equation}
with $R^{-1}_{\ell,k_0,k}$ as an element of the transformation matrix between the former basis element of index $k_0$, $p_{k_0}$ (before orthonormalisation), and the orthonormal basis element of index $k$, $q_k$, for a given spherical degree $\ell$. This definition corresponds to the slope of a linear regression of the frequencies as a function of the radial order \citep[similarly to][]{2012A&A...539A..63R} but expressed in the \who formalism.

FUrthermore, the helium glitch amplitude is expressed as
\begin{equation}\label{Fig:Who-AHe2}
\mathcal{A}_{\textrm{He}} = \frac{\sqrt{\sum\limits_k a_{\textrm{He},k}^2}}{\sqrt{\sum\limits_{i=1}^N 1/\sigma\left(\nu_i\right)^2}},
\end{equation}
with $a_{\textrm{He},k}$ the four projection coefficients on the basis elements associated to the helium glitch \citep[see Eq~20 of][]{2019A&A...622A..98F} and $\sigma\left(\nu_i\right)$ as the uncertainty associated with the i-th observed frequency. We recall that the glitch basis elements and projection coefficients are independent of the spherical degree per definition.

\subsection{Best model search, models and frequencies}\label{Sec:ModSrc}
We searched for models representative of a set of observed constraints using a Levenberg-marquardt algorithm, PORTE-CLES (Farnir \& Dupret 2025 in prep.). For each set of stellar parameters, $\mathbf{P}$, the quality of a given model is assessed through the following cost function:
\begin{equation}
    \chi^2\left(\mathbf{P}\right) = \sum^{K}_{i=1}\frac{\left(C_{i,\textrm{obs}}-C_{i,\textrm{mod}}\left(\mathbf{P}\right)\right)^2}{\sigma_i^2},\label{Eq:chi}
\end{equation}
with $C_i$ as the K constraints, either observed or modelled -- `obs' or `mod' subscripts respectively -- and $\sigma_i$ as the associated uncertainty.

\begin{figure}
    \centering
    \includegraphics[width=.95\linewidth]{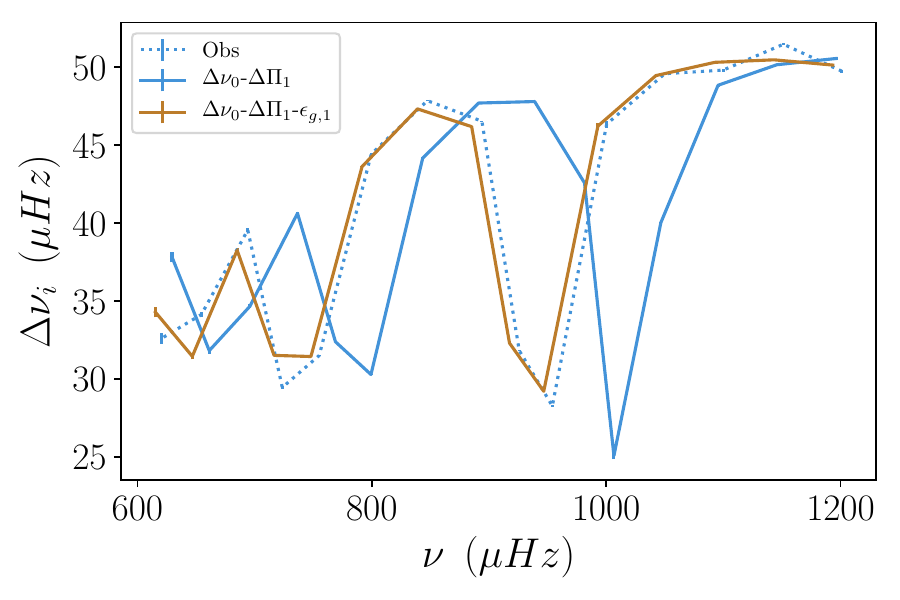}
    \caption{Comparison between observed pairwise frequency separations of dipolar modes and the modelled ones. The dotted blue line corresponds to the observation, the continuous blue line used only $\Delta\nu_0$ and $\Delta\Pi_1$ as fitting constraints, and the light brown one also includes the gravity offset $\epsilon_{g,1}$. All models are computed with an initial composition of $X_0=0.70$ and $Z_0=0.017$.}
    \label{Fig:DnuEg}
\end{figure}

Every model was computed using CLES \citep{2008Ap&SS.316...83S} with the standard physics presented in \cite{2019A&A...622A..98F} and the corresponding adiabatic oscillation frequencies were obtained using LOSC \citep{2008Ap&SS.316..149S}. The reference choices of input physics are recalled in Table~\ref{Tab:RefPhy}. Variations from these references will be explicitly stated in the text. Unless stated otherwise, model frequencies were not corrected for the surface effects.
With a fixed composition we searched for optimal models using the stellar age, $t$, and mass, $M$, as free parameters and $\Delta\nu_0$, $\Delta\Pi_1$, and $\epsilon_{g,1}$ as constraints. For each model produced by the optimisation procedure these indicators are computed from the model frequencies using \egg. We note that both $\Delta\Pi_1$ and $\epsilon_{g,1}$ are used simultaneously as constraints without including an additional free parameter, leading to an extra constraint compared to the number of free parameters. Indeed, it can be expected from Eq.~\ref{Eq:tg} and as was noted by \citet[][{\href{https://www.aanda.org/articles/aa/pdf/2021/09/aa41317-21.pdf\#figure.4}{Fig.~4}}]{2021A&A...653A.126F} that both constraints are highly correlated. A small change of $\Delta\Pi_1$ amounts to changing the slope of the linear relation expected between modes' periods and their radial order. Consequently, the best fit intercept, corresponding to $\epsilon_{g,1}$, is significantly modified. It is therefore necessary to adjust those two constraints simultaneously. Figure~\ref{Fig:DnuEg} illustrates the impact of including $\epsilon_{g,1}$ in the set of constraints. It displays pairwise frequency separations, $\Delta\nu_i$, between successive dipolar oscillation modes. We observe, when comparing the observed data (dashed light blue curve) and the model one with only $\Delta\nu_0$ and $\Delta\Pi_1$ that the separation between the dips left by the g-dominated mixed modes is well accounted for. This is a result of the good agreement with the period spacing value. The same goes for the height of the bumps, which corresponds to the large frequency separation. However, there is a clear shift between the two curves. The inclusion of the gravity offset, $\epsilon_{g,1}$, allows to account for this shift as shown by the light brown curve. We add that, while the improvement is obvious, the changes in optimal parameters are negligible (with $\epsilon_{g,1}$: $M=1.168~M_{\odot}$, $t=5.96~\textrm{Gyr}$; and $M=1.169~M_{\odot}$, $t=5.93~\textrm{Gyr}$ without). The main change occurs in age and amounts for a relative variation of about $0.5\%$, which is sufficient to produce the necessary phase shift for the two curves to match at best.

\begin{table}[h]
    \centering
    \caption{Reference stellar parameters used in our models.}
    \begin{tabular}{c|c}
    Input physics & Choice \\
    \hline
    \hline
        Metal mixture & 1 \\
        Opacity (high T) & 2 \\
        Opacity (low T) & 3 \\
        Equation of state & 4 \\
        Nuclear reaction rates & 5 \\
        Mixing length parameter, $\alpha_{\textrm{MLT}}$ & 1.80 (solar calibration) \\
        Overshooting parameter, $\alpha_{\textrm{ov}}$ & 0.00 \\
        Initial hydrogen mass fraction, $X_0$ & $0.72$ \\
        Initial metals mass fraction, $Z_0$ & $0.015$ \\
    \end{tabular}
    \tablebib{(1)~\citet{2009ARA&A..47..481A}; (2)~\citet{1996ApJ...464..943I}; (3)~\citet{2005ApJ...623..585F}; (4)~\citet{2003ApJ...588..862C}; (5)~\citet{2011RvMP...83..195A}}
    \label{Tab:RefPhy}
\end{table}

\section{Results}
In this section, we present and discuss the results of our modelling (all displayed in Table~\ref{Tab:AllRes}). We do not provide the uncertainties on the two free parameters as they are the result of error propagation returned by the minimisation procedure which are unrealistically small: $0.2\cdot 10^{-3}~\textrm{Gyr}$ in age and $0.1\cdot 10^{-4}~M_{\odot}$ in mass. The uncertainties determination is discussed in Sect.~\ref{Sec:Dis}. We distinguish our results according to the set of input physics we tested. First, in Sect.~\ref{Sec:Comp}, we consider an ensemble of chemical compositions and assess its impact on the optimal parameters as well as on the auxiliary constraints. Then, we study the inclusion of overshooting and the impact of varying the mixing-length parameter in Sect.~\ref{Sec:Mix}. We also characterise the information carried by quadrupolar modes and the possibility to use overshooting to account for the composition profile in H-shell in Sect.~\ref{Sec:Quad}. Finally, we study the variation in optimal parameters when correcting for the surface effects as prescribed in \citet{2014A&A...568A.123B} in Sect.~\ref{Sec:Surf}.

\begin{table*}[]
    \centering
    \caption{Summary of all computed models.}
    \begin{tabular}{c*{11}{|c}}
         $X_0$ & $Z_0$ & $\alpha_{\textrm{ov}}$ & $\alpha_{\textrm{MLT}}$ & Surf. & $M~\left(M_{\odot}\right)$ & $t~\left(\textrm{Gyr}\right)$ & $\rho_c~\left(\textrm{g}/\textrm{cm}^3\right)$ & $\overline{\rho}~\left(\textrm{g}/\textrm{cm}^3\right)$ & $\rho_c/\overline{\rho}$ & $L\left(L_\odot\right)$ & $R\left(R_\odot\right)$ \\
         \hline
         \hline
         $0.70$ & $0.015$ & $0.00$ & $1.80$ & none & $1.15$ & $5.91$ & $3.88\cdot 10^3$ & $1.86\cdot 10^{-1}$ & $2.08\cdot 10^4$ & $4.05$ & $2.05$ \\
         $0.70$ & $0.017$ & $0.00$ & $1.80$ & none & $1.17$ & $5.96$ & $3.88\cdot 10^3$ & $1.87\cdot 10^{-1}$ & $2.08\cdot 10^4$ & $3.96$ & $2.07$ \\
         $0.72$ & $0.013$ & $0.00$ & $1.80$ & none & $1.16$ & $6.03$ & $3.88\cdot 10^3$ & $1.86\cdot 10^{-1}$ & $2.08\cdot 10^4$ & $4.17$ & $2.06$ \\
         $0.72$ & $0.015$ & $0.00$ & $1.80$ & none & $1.18$ & $6.08$ & $3.88\cdot 10^3$ & $1.86\cdot 10^{-1}$ & $2.08\cdot 10^4$ & $4.05$ & $2.07$ \\
         $0.72$ & $0.017$ & $0.00$ & $1.80$ & none & $1.20$ & $6.15$ & $3.89\cdot 10^3$ & $1.87\cdot 10^{-1}$ & $2.09\cdot 10^4$ & $3.97$ & $2.09$ \\
         $0.74$ & $0.015$ & $0.00$ & $1.80$ & none & $1.21$ & $6.27$ & $3.89\cdot 10^3$ & $1.86\cdot 10^{-1}$ & $2.08\cdot 10^4$ & $4.07$ & $2.09$ \\
         \hline
         $0.70$ & $0.015$ & $0.15$ & $1.80$ & none & $1.18$ & $5.26$ & $3.86\cdot 10^3$ & $1.86\cdot 10^{-1}$ & $2.07\cdot 10^4$ & $4.47$ & $2.08$ \\
         $0.70$ & $0.017$ & $0.15$ & $1.80$ & none & $1.20$ & $5.28$ & $3.87\cdot 10^3$ & $1.87\cdot 10^{-1}$ & $2.07\cdot 10^4$ & $4.36$ & $2.09$ \\
         $0.72$ & $0.013$ & $0.15$ & $1.80$ & none & $1.20$ & $5.31$ & $3.85\cdot 10^3$ & $1.87\cdot 10^{-1}$ & $2.06\cdot 10^4$ & $4.62$ & $2.08$ \\
         $0.72$ & $0.015$ & $0.15$ & $1.80$ & none & $1.22$ & $5.32$ & $3.86\cdot 10^3$ & $1.86\cdot 10^{-1}$ & $2.07\cdot 10^4$ & $4.50$ & $2.10$ \\
         $0.72$ & $0.017$ & $0.15$ & $1.80$ & none & $1.24$ & $5.41$ & $3.86\cdot 10^3$ & $1.86\cdot 10^{-1}$ & $2.07\cdot 10^4$ & $4.39$ & $2.11$ \\
         $0.74$ & $0.015$ & $0.15$ & $1.80$ & none & $1.26$ & $5.46$ & $3.86\cdot 10^3$ & $1.87\cdot 10^{-1}$ & $2.07\cdot 10^4$ & $4.53$ & $2.12$ \\
         \hline
         $0.70$ & $0.017$ & $0.00$ & $1.73$ & none & $1.15$ & $6.29$ & $3.89\cdot 10^3$ & $1.86\cdot 10^{-1}$ & $2.09\cdot 10^4$ & $3.75$ & $2.06$ \\
         $0.70$ & $0.017$ & $0.15$ & $1.73$ & none & $1.18$ & $5.73$ & $3.96\cdot 10^3$ & $1.90\cdot 10^{-1}$ & $2.08\cdot 10^4$ & $4.03$ & $2.06$ \\
         \hline
         $0.70$ & $0.017$ & $0.17$ & $1.80$ & none & $1.20$ & $5.51$ & $3.87\cdot 10^3$ & $1.86\cdot 10^{-1}$ & $2.07\cdot 10^4$ & $4.25$ & $2.08$ \\
         \hline
         $0.70$ & $0.017$ & $0.00$ & $1.80$ & BG14 & $1.17$ & $5.97$ & $3.89\cdot 10^3$ & $1.86\cdot 10^{-1}$ & $2.08\cdot 10^4$ & $3.96$ & $2.07$ \\
    \end{tabular}
    \tablefoot{Columns $1$ to $5$ correspond to fixed parameters in the model search, columns $7$ and $8$ are the optimised free parameters, and columns $9$ to $11$ are derived quantities from the models. In the order of the table, we have: the initial hydrogen mass fraction, the initial metals fraction, the overshooting parameter, the mixing-length parameter, the surface effects prescription (either none or \cite{2014A&A...568A.123B}'s `inverse-cubic'), the optimised mass, the optimised age, the central density, the mean density and the central to mean density ratio. We have separated the different cases considered throughout the paper by a horizontal line.}
    \label{Tab:AllRes}
\end{table*}

\subsection{Impact of the composition}\label{Sec:Comp}
\begin{figure}
    \centering
    \includegraphics[width=.95\linewidth]{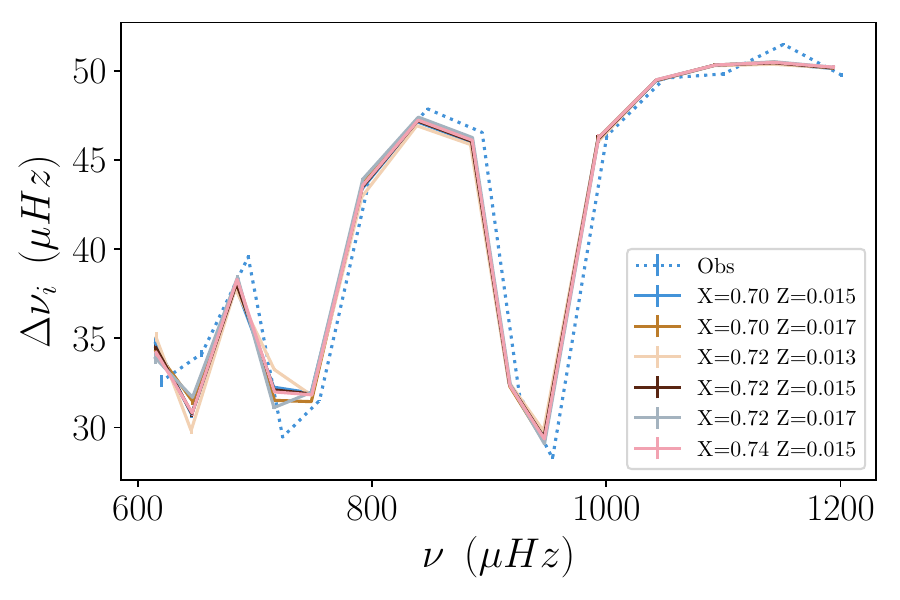}
    \caption{Comparison between observed pairwise frequency separations of dipolar modes and the modelled ones, without overshooting. The dotted pale blue curve correspond to the observed values while the continuous ones correspond to the variations in initial chemical compositions.}
    \label{Fig:DnuNoOver}
\end{figure}

\begin{figure}
    \centering
    \includegraphics[width=.95\linewidth]{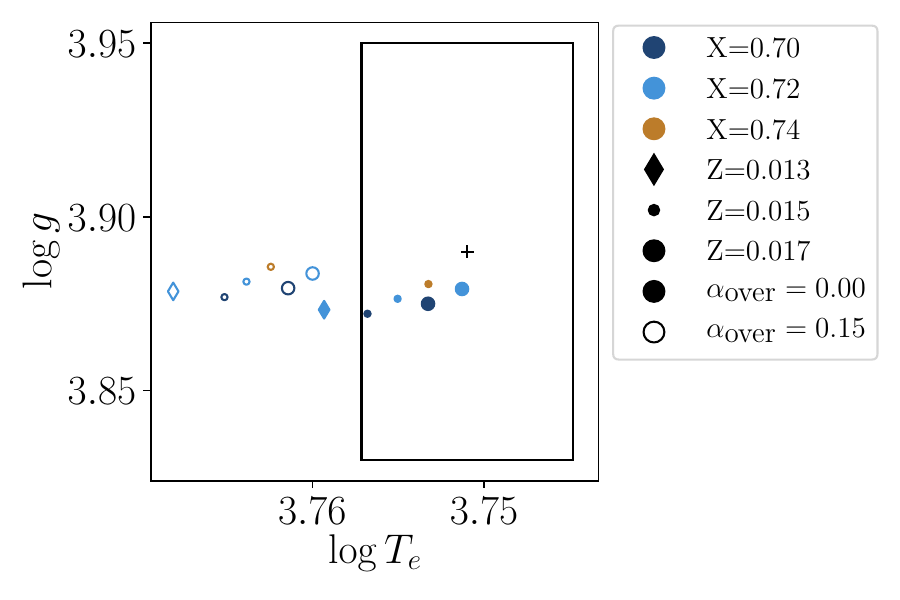}
    \caption{Kiel diagram of the optimal models compared with the observed values, black box. Colour variations correspond to variations in initial hydrogen mass fraction and symbols to variations in initial metals fraction. Filled symbols are for models without overshooting while the empty ones have an overshooting parameter of $\alpha_{\textrm{ov}}=0.15$.}
    \label{Fig:Kiel}
\end{figure}

\begin{figure}
    \centering
    \includegraphics[width=.95\linewidth]{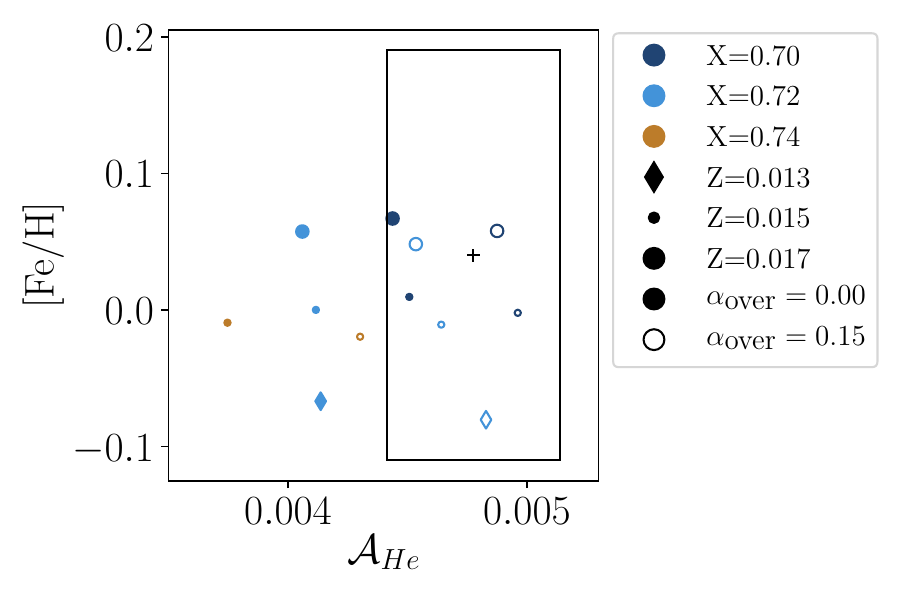}
    \caption{Surface metallicity and helium glitch amplitude of optimal models compared with the observed values. The legend is the same as in Fig.~\ref{Fig:Kiel}.}
    \label{Fig:Comp}
\end{figure}

With fixed input physics, along with mixing-length and overshooting parameters, we built an ensemble of models following the methodology described in Sect.~\ref{Sec:ModSrc} for different compositions. Because Gemma's metallicity is compatible with that of our Sun \citep[${\left[\textrm{Fe/H}\right]}=0.04 \pm 0.15$, ][]{2012A&A...543A..54A}, here we consider compositions centered around values typical of the Sun; namely, $X_0=0.72$ and $Z_0=0.015$, obtained via a solar calibration. We evaluated the impact of changes of $0.02$ in $X_0$ and $0.002$ in $Z_0$ on the optimal models retrieved (i.e. minimising the cost function given in Eq.~\ref{Eq:chi}). The results are displayed in Fig.~\ref{Fig:DnuNoOver} representing individual dipolar frequency separation for each model (coloured continuous lines) compared with the observed values (blue dotted line). Ordering modes of the same spherical degree, i.e. $\ell=1$, in increasing order, these frequency separations correspond to the difference between two successive modes. We observe that the agreement between the observed differences and all the model ones is excellent. Nonetheless, from Fig.~\ref{Fig:DnuNoOver} alone, it is difficult and somewhat arbitrary to select one composition over the others. Indeed, a $\chi^2$ quantification of the agreement between observed and theoretical frequency differences yields similar results for all the models\footnote{We recall that the cost function used in the model search does not explicitly include individual frequencies but rather seismic indicators that are combinations of these frequencies.}. Therefore, we evaluated the models agreement with additional constraints, namely $\log g$, $T_{\textrm{eff}}$, $\left[\textrm{Fe/H}\right]$, and the helium glitch amplitude $\mathcal{A}_{\textrm{He}}$. This is represented in Figs.~\ref{Fig:Kiel} and \ref{Fig:Comp} where the colours of the symbols represent the initial hydrogen mass fraction, their shapes show the initial metals mass fraction and filled symbols show the models without overshooting while empty symbols include overshooting (as discussed in subsequent sections). From these figures, it becomes immediately apparent that not all models are built equal. First, we note that the spectroscopic $\log g$ adds no information when used concurrently with seismic data. However, while most models fall within the observed uncertainties for $T_{\textrm{eff}}$, the model with a reduced amount of metals ($X_0=0.72$ and $Z_0=0.013$) does not agree with the observations. This shows the necessity to account for this constraint in our modelling. In addition, we note that the surface metallicity and helium glitch amplitude also provide relevant pieces of information. Indeed, the models centered around the reference composition ($X_0=0.72$ and $Z_0=0.015$) all display a helium glitch amplitude that is too low when compared with the observations. \cite{2004MNRAS.350..277B,2019MNRAS.483.4678V,2019A&A...622A..98F} have shown that the $\mathcal{A}_{\textrm{He}}$ correlates with the amount of helium in the envelope. Therefore, increasing the initial helium abundance, $Y_0$, by reducing $X_0$ should help. This can be expected from the relative displacement in Fig.~\ref{Fig:Comp} when changing only $X_0$ (same symbol, different colours). By trying out several combinations, we find that a model with low hydrogen and high metal content ($X_0=0.70$, $Z_0=0.017$) is able to reproduce the observed helium glitch amplitude (noting that the spectroscopic metallicity uncertainties are too large to be discriminating). This favours a helium rich model as well with $Y_0=0.283$. While these constraints were only considered in post-treatment of our results, this hints at the possibility to lift the well-established helium-mass degeneracy \citep{2014A&A...569A..21L}. We also note the surprising decrease in amplitude for the $X_0=0.70$ - $Z_0=0.015$ composition relative to the one reproducing the helium glitch amplitude, even though it is initially richer in helium ($Y_0=0.285$). We must recall that, while at fixed stellar parameters, an increase in initial helium abundance means a stronger helium glitch signature, there is no direct relation between these quantities when adjusting several parameters and the final age and mass also play a role. Indeed, those two models display notable differences in these fundamental parameters. As noted by several authors \citep[e.g.][]{2004MNRAS.350..277B,2019A&A...622A..98F}, the helium glitch amplitude is correlated with the stellar mass.

\subsection{Constraining mixing processes}\label{Sec:Mix}
\subsubsection{Convective overshooting}\label{Sec:Ov}
\begin{figure}
    \centering
    \includegraphics[width=.95\linewidth]{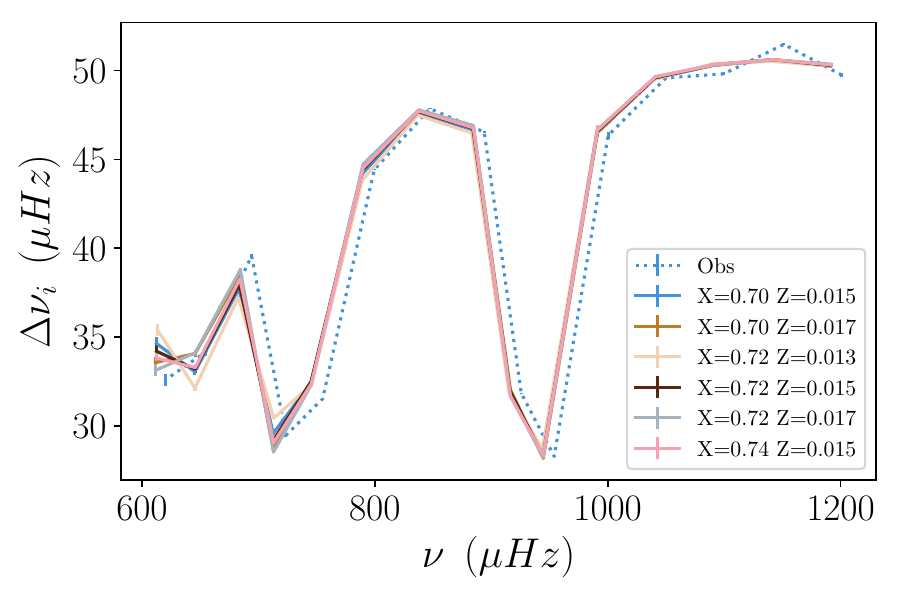}
    \caption{Comparison between observed pairwise frequency separation and the modelled ones, with an overshooting parameter of $\alpha_{\textrm{ov}}=0.15$. The legend is the same as in Fig.~\ref{Fig:DnuNoOver}.}
    \label{Fig:DnuOver}
\end{figure}

  To go further, we tested the impact of core overshooting on the optimal results. To do so, we included step overshooting with the temperature gradient set to the radiative one\footnote{This prescription for overshooting simply amounts to extending the mixed region of the convective core beyond its classical boundary dictated by the Ledoux criterion.}, with $\alpha_{\textrm{ov}}=0.15$, in our set of models, with the same compositions as in the previous section. We show the frequency differences for these models in Fig.~\ref{Fig:DnuOver}. Visually, the results seem to have improved. This is confirmed by a $\chi^2$ analysis comparing the observed and theoretical frequency separations -- models without overshooting typically have $\chi^2 \sim 1000$, while those with $\alpha_{\textrm{ov}}=0.15$ yield $\chi^2 \sim 100$. This tends to favour models with overshooting over those without. Nevertheless, we also placed these models in the Kiel and the metallicity - helium glitch amplitude diagrams from earlier sections of this work (Figs.~\ref{Fig:Kiel} and \ref{Fig:Comp}). From these figures, we may draw somewhat different conclusions. Indeed, we observe that none of the models with overshooting reproduce the observed effective temperature. The inclusion of overshooting leads to a systematic increase in temperature of about $130~\textrm{K}$, which greatly exceeds the observed uncertainties of $80~\textrm{K}$. However, the effective temperature alone does not serve as a sufficient criterion to exclude certain models as its determination is somewhat uncertain. Indeed, literature values range from $5487 \pm 69~\textrm{K}$ \citep{2017ApJS..233...23S} to $5802 \pm 68~\textrm{K}$ \citep{2013MNRAS.434.1422M} which is compatible with most models. In addition, the $\left[\textrm{Fe/H}\right]$-$\mathcal{A}_{\textrm{He}}$ diagram provides us with yet another picture. We observe that most models agree with the composition constraints with the exception of the helium-poorest one ($X_0=0.74$ - $Z_0=0.015$), which was already in disagreement, even without overshooting.

\begin{figure}
    \centering
    \includegraphics[width=.95\linewidth]{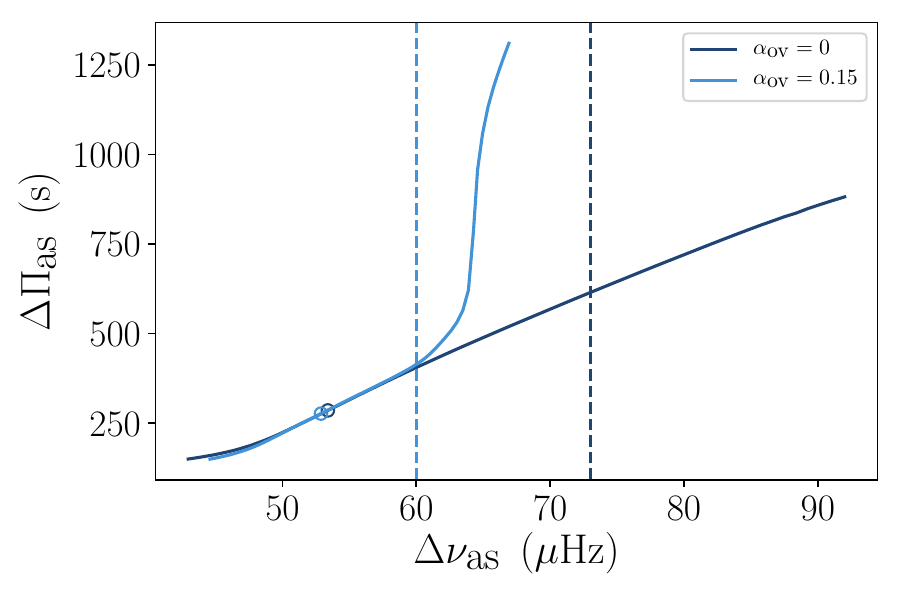}
    \caption{Evolution (from right to left) of $\Delta\Pi_{\textrm{as}}$ as a function of $\Delta\nu_{\textrm{as}}$ for models with an initial composition of $X_0=0.70$ and $Z_0=0.017$ either including overshooting ($\alpha_{\textrm{ov}}=0.15$, light blue) or not (dark blue). The evolution, from right to left, starts at the late main-sequence phase and evolves past the beginning of the subgiant phase (symbolised by the vertical dashed lines). The best fit models for each prescription are shown with the empty circles.}
    \label{Fig:DnuDpi}
\end{figure}

While none of our models retained a convective core on the subgiant phase, there are notable differences between the two prescriptions for the overshooting. These differences are fossil signatures left by the MS convective core overshooting. This is illustrated in Fig.~\ref{Fig:DnuDpi}, which traces the evolution of the asymptotic values (for smoothness) of the large frequency separation, $\Delta\nu_{\textrm{as}}$, and dipolar period spacing, $\Delta\Pi_{\textrm{as}}$, (Eqs.~\ref{Eq:Dnu} and \ref{Eq:Dpi}) from the late MS phase to the subgiant one. The approximate beginning of the subgiant phase is indicated by the vertical dashed lines. Then $\Delta\nu_{\textrm{as}}$ and $\Delta\Pi_{\textrm{as}}$ are defined as:

\begin{equation}
    \Delta\nu_{\textrm{as}} = \int_{\textrm{p-cavity}} \frac{dr}{c},\label{Eq:Dnu}
\end{equation}
and
\begin{equation}
    \Delta\Pi_{\textrm{as}} = 2\pi^2\left(\int_{\textrm{g-cavity}} \frac{N}{r}dr\right)^{-1},\label{Eq:Dpi}
\end{equation}
with $c$ the local sound speed value and $N$ the Brunt-Väisälä frequency (Eq.~\ref{Eq:Brunt}). The p-cavity (resp. g-cavity) denotes the region of propagation of p-modes (resp. g-modes).

From Fig.~\ref{Fig:DnuDpi}, it is clear that the two sequences of evolution are significantly different on the MS. This is due to the presence (resp. absence) of a convective core with overshooting (resp. without). Indeed, the convective core reduces the size of the g-cavity ($N=0$ in convective cores), which leads to a smaller integral in Eq.~\ref{Eq:Dpi} and, therefore a larger period spacing. Then, in the subgiant phase, symbolised by the vertical dashed lines, the two curves converge to a common evolution. This shows that there subsist small nuances in the oscillation spectra that are not captured in the asymptotic $\Delta\nu_{\textrm{as}}$ and $\Delta\Pi_{\textrm{as}}$. Indeed, $\epsilon_{g,1}$ was also part of the constraints of the adjustments for every model. This suggests that the convective core leaves a signature that introduces a subtle shift in phase of the oscillations. The inclusion of the $\epsilon_{g,1}$ constraint also justifies that the two optimal models, represented as the empty circles, do not perfectly line up. Indeed, doing so, we have an extra constraint compared to the number of free parameters.

\begin{figure}
    \centering
    \includegraphics[width=.95\linewidth]{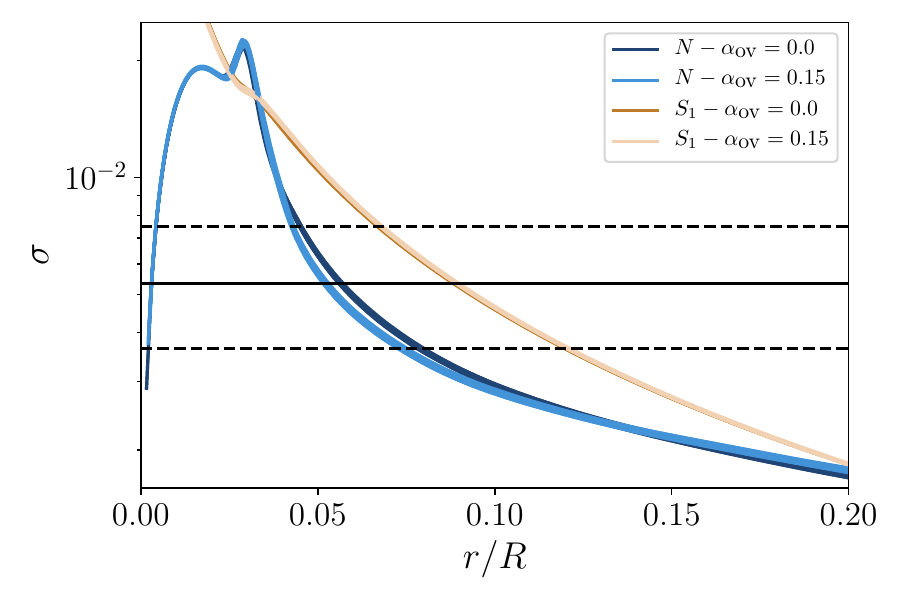}
    \caption{Propagation diagram for all of the optimal models presented in Sect.~\ref{Sec:Ov}. The Brunt-Väisälä frequency of the models without overshooting (resp. with $\alpha_{\textrm{ov}}=0.15$) is represented as a function of the reduced radius in dark blue (resp. light blue). The Lamb frequency of dipolar modes is shown in brown (resp. beige) for models without (resp. with) overshooting. The black continuous line corresponds to the frequency at maximum power and the dashed black lines correspond to the observed frequency range.}
    \label{Fig:Prop}
\end{figure}

\begin{figure}
    \centering
    \includegraphics[width=.95\linewidth]{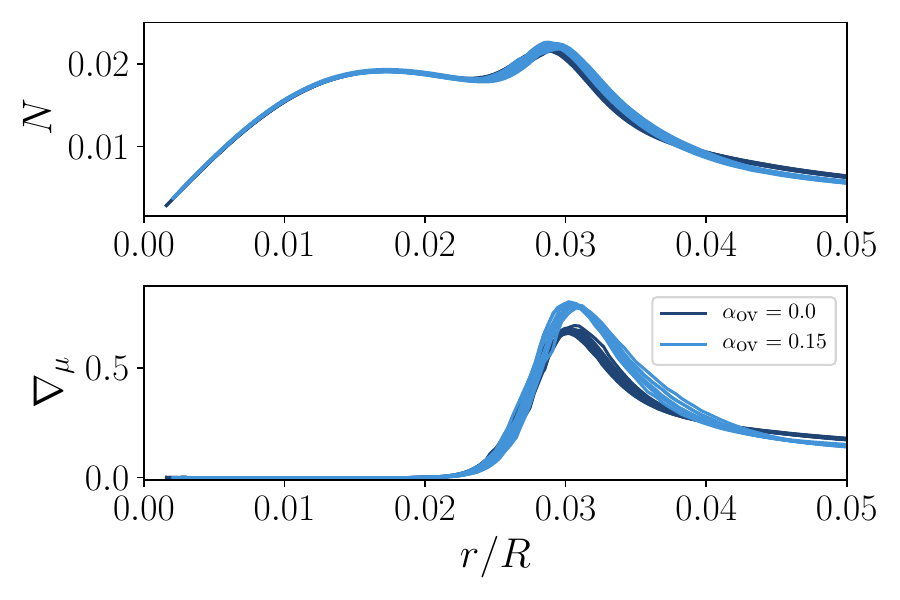}
    \caption{\textbf{top panel}: Same Brunt-Väisälä frequency profiles as in Fig.~\ref{Fig:Prop} from the center to the H-shell. \textbf{bottom panel}: Mean molecular weight of the optimal models over the same radius range as the upper panel. In both panels, dark blue corresponds to models without overshooting while light blue curves correspond to those with overshooting.}
    \label{Fig:mu}
\end{figure}

\begin{figure}
    \centering
    \includegraphics[width=.95\linewidth]{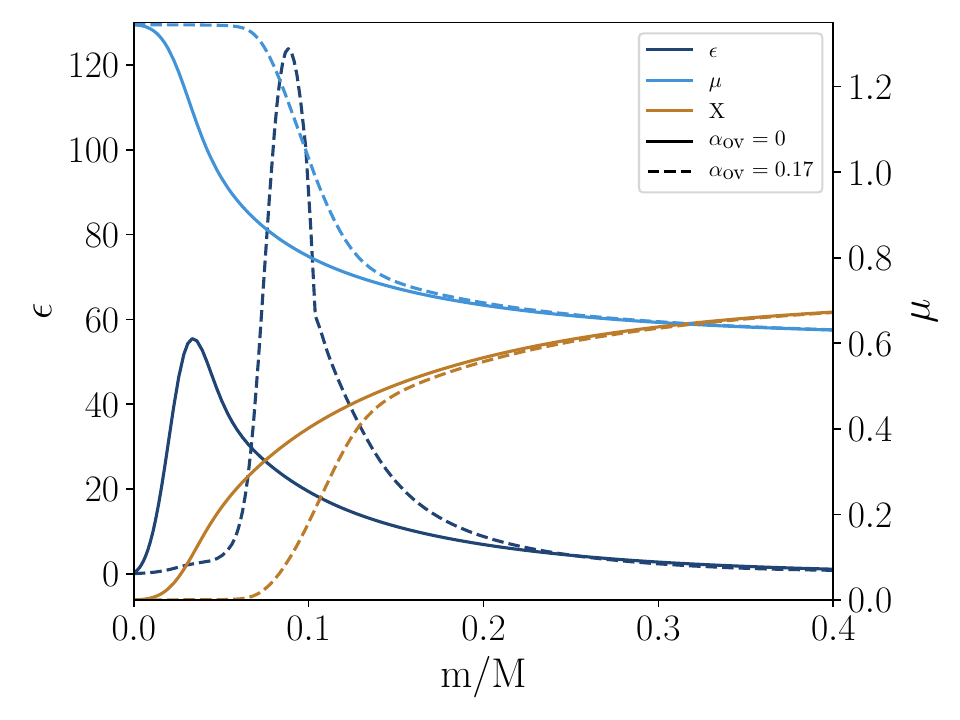}
    \caption{Comparison of the energy produced by nuclear reactions (in $\textrm{erg}~g^{-1}s^{-1}$, dark blue) as a function of the reduced mass for two stars at the TAMS (set at $X_c = 10^{-4}$) for models with the best composition ($X_0=0.70$ - $Z_0=0.017$) with (dashed lines) and without overshooting (continuous lines). We also display the mean molecular weight (light blue) and hydrogen (light brown) profiles.}
    \label{Fig:nucTAMS}
\end{figure}

\begin{figure}
    \centering
    \includegraphics[width=.95\linewidth]{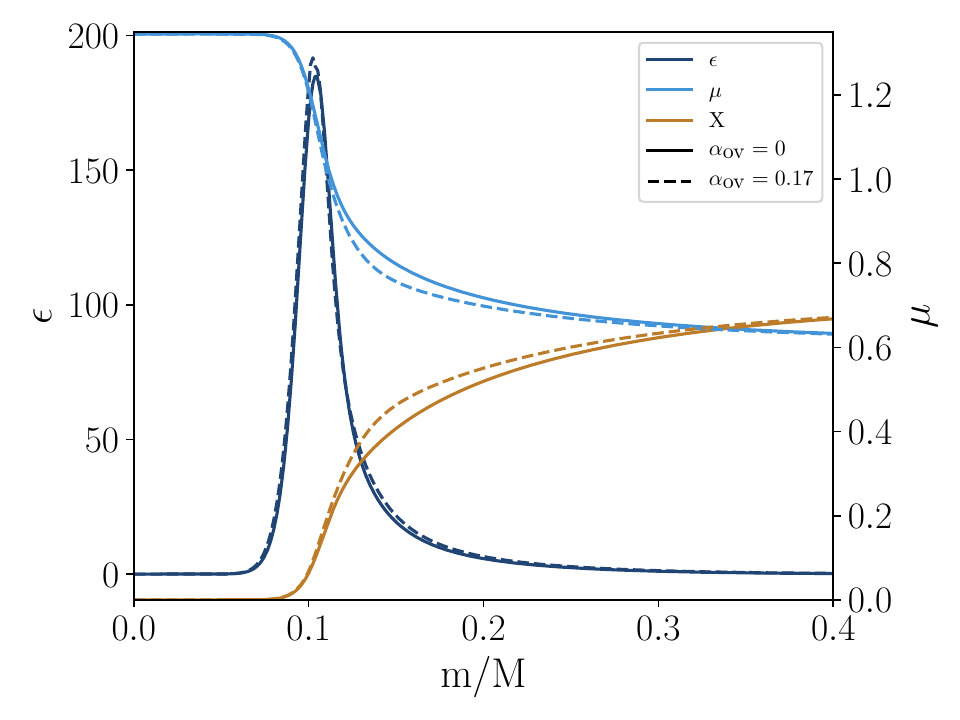}
    \caption{Comparison of the energy produced by nuclear reactions (in $\textrm{erg}~g^{-1}s^{-1}$, dark blue) as a function of the reduced mass for the two fully evolved optimal models with the best composition ($X_0=0.70$ - $Z_0=0.017$) with (dashed lines) and without overshooting (continuous lines). We also display the mean molecular weight (light blue) and hydrogen (light brown) profiles.}
    \label{Fig:nucOpt}
\end{figure}

To further interpret the influence of overshooting on our results, we display the propagation diagram of all of the optimal models with and without overshooting in the innermost regions of the star in Fig.~\ref{Fig:Prop}. In dark blue (respectively light blue) is represented the Brunt-Väisälä frequency, $N$, as a function of the reduced radius of the models without overshooting (resp. with $\alpha_{\textrm{ov}}=0.15$). We show the Lamb frequency of dipolar modes in brown (respectively beige) for models without overshooting (resp. with $\alpha_{\textrm{ov}}=0.15$). Also, the frequency of maximum power, $\nu_{\textrm{max}}$, is displayed as the continuous black line and the observed range as the dashed lines. First, we note that curves with the same prescription for the overshooting line up almost perfectly, forming two distinct families. This is reassuring as it demonstrates the stability of our approach towards a preferred structure. Then, we note that the main changes happen in the evanescent region (between the Lamb and Brunt-Väisälä frequencies ($r/R \in \sim \left[0.05,0.075 \right]$). In this region the two $N$ frequencies show distinct profiles. As a matter of fact, the observed differences are a direct consequence of variations in the composition profile in the H-shell, corresponding to the bump in $N$ as can be expected from its expression (see Eq.\ref{Eq:Brunt}).
Consequently, we display, in Fig.~\ref{Fig:mu}, $N$ from the center to the H-shell (upper panel) as well as $\nabla_\mu$ (lower panel). We observe that the two composition gradient profiles are similar in the innermost region and that they only separate in the shell, where the gradient is steeper (higher value) for the models with overshooting (light blue) compared to those without overshooting (dark blue). We also observe that the H-shell is narrower for models with overshooting when compared to those without. This is due to the recession of the convective core of models with overshooting (while those without are not able to retain a convective core on the MS) that leaves a steep composition gradient at the boundary of the shell at the end of the MS. To illustrate these structural differences at the boundary of the convective core, we computed the complete evolutionary sequences with the initial parameters of two optimal models, with an overshooting parameter of $\alpha_{\textrm{ov}}=0.17$\footnote{A justification for this specific value will be provided in Sect.~\ref{Sec:Quad}.} and without overshooting, with an initial composition of $X_0=0.70$ and $Z_0=0.017$. We show the mean molecular weight, hydrogen and total nuclear energy generation profiles as a function of reduced mass at the terminal age main sequence (TAMS) in Fig.~\ref{Fig:nucTAMS} and for the optimal model (i.e. reproducing at best the observed $\Delta\nu_0$, $\Delta\Pi_1$, and $\epsilon_{g,1}$) in Fig.~\ref{Fig:nucOpt}. First, we note at the TAMS that the shell, corresponding to the nuclear energy production peak, lies in `shallower' regions in the model with overshooting than without. The resulting is a steeper composition gradient in models with overshooting. As both models finish their evolution, these structural differences are mostly erased. Nonetheless, we observe that the composition gradient remains steeper for models with overshooting. Therefore, we expect that what can be probed via asteroseismology is this composition gradient and that overshooting is one of the possible mechanisms that can account for such a profile. We add that, while it would seem surprising that composition gradient in models without overshooting would be able to catch up with that of models with overshooting, this is merely an effect of time alone. Indeed, the duration of the transition between the TAMS and optimal model (in the subgiant phase) takes vastly different amounts of time in both cases as the main-sequence phase do not span the same duration. The models with overshooting need only $\sim0.15~\textrm{Gyr}$, while those without evolve for about $\sim1.25~\textrm{Gyr}$, allowing for the two profiles to ultimately be matched.

\subsubsection{Mixing length}
\begin{figure}
    \centering
    \includegraphics[width=.95\linewidth]{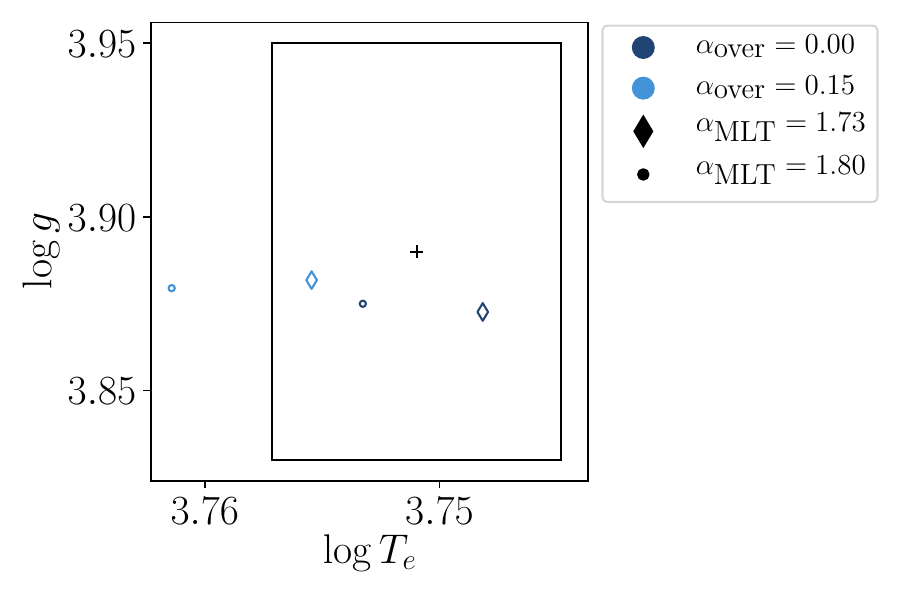}
    \caption{Kiel diagram of the optimal models compared with the observed values, black box. Colour variations correspond to variations in initial hydrogen mass fraction and symbols to variations in initial metals fraction. Filled symbols are for models without overshooting while the empty ones have an overshooting parameter of $\alpha_{\textrm{ov}}=0.15$.}
    \label{Fig:Kiel-MLT}
\end{figure}

\begin{figure}
    \centering
    \includegraphics[width=.95\linewidth]{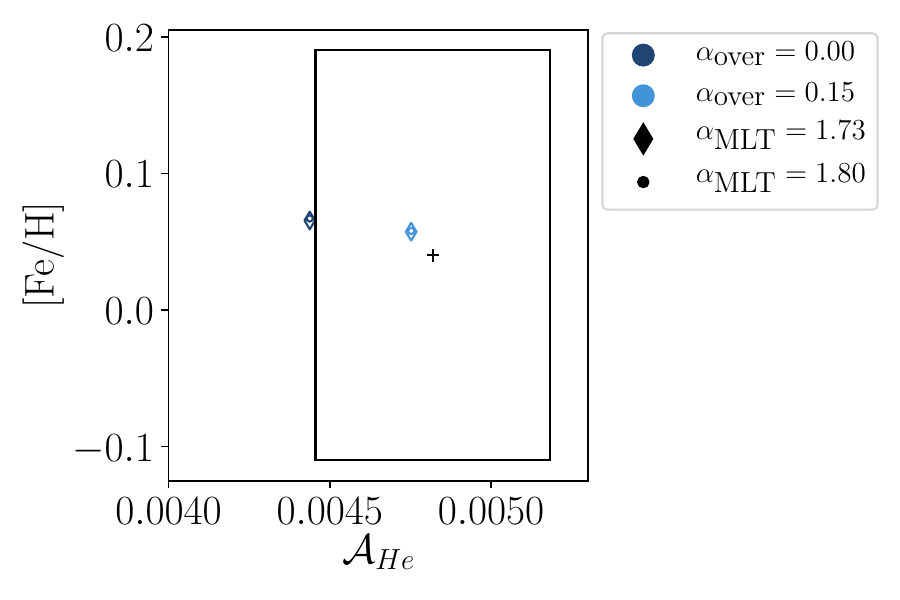}
    \caption{Surface metallicity and helium glitch amplitude of optimal models compared with the observed values. The legend holds the same meaning as in Fig.~\ref{Fig:Kiel-MLT}.}
    \label{Fig:Comp-MLT}
\end{figure}

To further characterise the impact of the convection modelling on our results, we computed models using a different value for the mixing-length parameter. To do so, we rely on the study of \cite{2015A&A...573A..89M} to inform our value of choice. They demonstrated through hydrodynamic simulations that the mixing-length parameter should vary with stellar surface properties such as the metallicity, effective temperature and surface gravity. We used their functional fit, presented in appendix B, and estimated that we ought to decrease the mixing-length parameter by about $0.07$ relative to our solar reference. Thus, we obtained a revised value of $\alpha_{\textrm{MLT}}=1.73$,  which we used to compute additional models (with $X_0=0.70$ and $Z_0=0.017$, favoured by the composition constraints), with and without overshooting. The resulting diagrams are shown in Figs.~\ref{Fig:Kiel-MLT} and \ref{Fig:Comp-MLT}, where the colours now correspond to the overshooting parameter and the symbols to the mixing-length one. We observe that there is next to no impact on the surface composition as all the markers with the same colour almost perfectly superimpose on one another. It is only the overshooting parameter, as observed earlier, that has a visible impact. Now looking at Fig.~\ref{Fig:Kiel-MLT}, we note that decreasing the mixing-length parameter value tends to decrease the effective temperature at an almost constant surface gravity. This is similar but in the opposite direction to the increase in effective temperature caused by the increase in overshooting. This interplay suggests that, in some cases, it might be possible and necessary to modify both parameters simultaneously to reproduce both seismic and non-seismic data.

\subsection{Information carried by the quadrupolar mixed modes}\label{Sec:Quad}
\begin{figure}
    \centering
    \includegraphics[width=.95\linewidth]{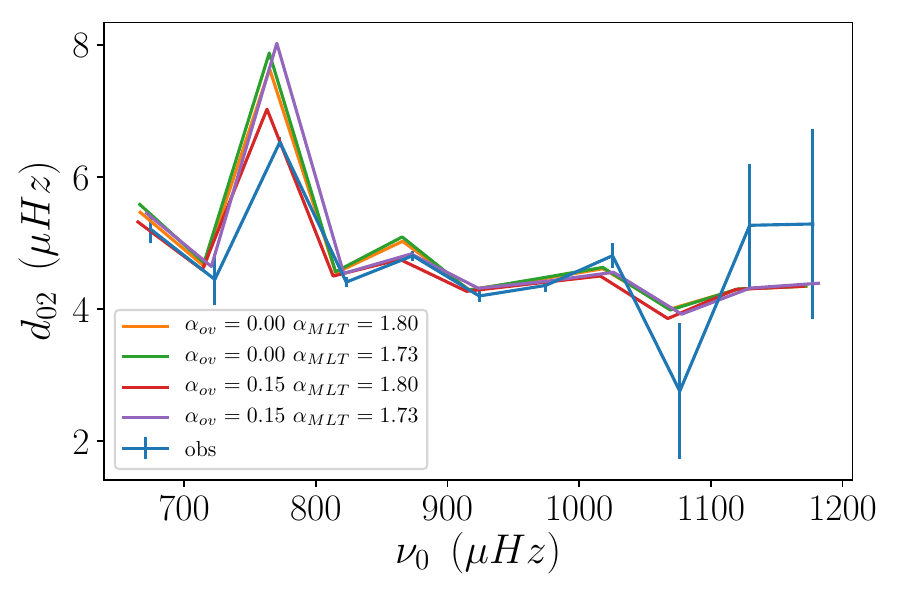}
    \caption{Individual small separation between quadrupolar and radial modes as a function of the radial frequency for different mixing-length and overshooting parameters (as detailed in the legend).}
    \label{Fig:d02}
\end{figure}

\begin{figure}
    \centering
    \includegraphics[width=.95\linewidth]{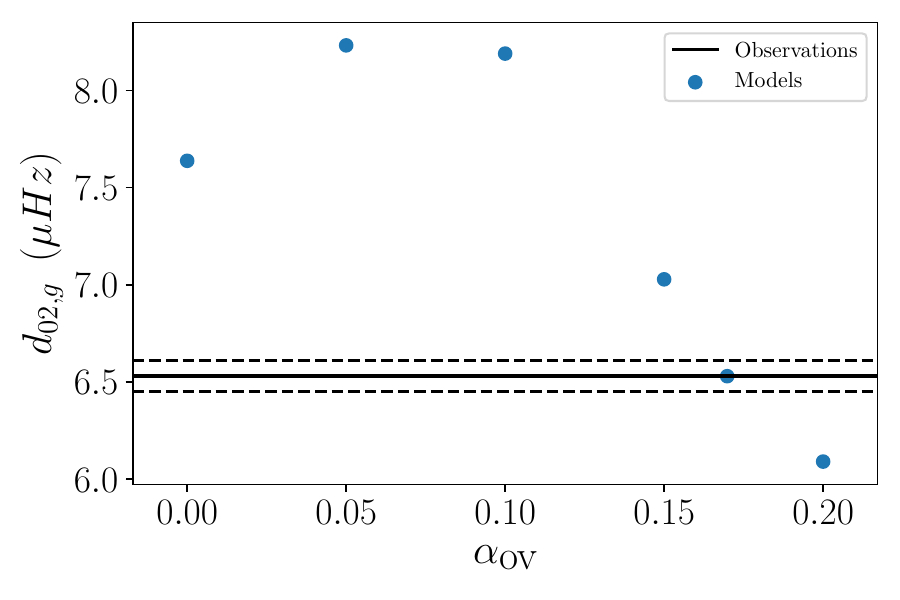}
    \caption{Evolution of the small separation of the quadrupolar small separation, $d_{02,\textrm{g}}$, with the overshooting parameter. The black continuous line corresponds to the observed value while the dashed ones are the associated uncertainties. Each blue point is a model adjusting $\Delta\nu_0$, $\Delta\Pi_1$, and $\epsilon_{g,1}$.}
    \label{Fig:l2MM}
\end{figure}

To further constrain our models, we characterise the information carried by the $\ell=2$ modes. First, we assess whether individual small separations between radial and quadrupolar modes may provide additional insight. They are defined as:
\begin{equation}
    \delta_{02,n} = \nu_{0,n}-\nu_{2,n-1},\label{Eq:d02}
\end{equation}
where $n$ represents the radial order of modes. We point out that the identification of mixed modes is a difficult task. Nonetheless, systematically pairing a radial mode with the closest quadrupolar one allows us to compute such small separations. These are represented in Fig.~\ref{Fig:d02} as a function of the radial mode frequency used to calculate them. We show the observed data and their respective uncertainties, in blue, as well as four models considering four combinations of overshooting and mixing-length parameter. First, we note that the last three modes display uncertainties that are too large for them to be used reliably. Then, aside the mode around $756~\mu Hz$ that displays a peak, the differences remain very similar to one another. This peak is actually caused by the presence of an $\ell=2$ g-dominated mixed mode, while the closest $\ell=2$ p-mode is absent in the observations, `artificially' increasing the small separation. We focus on this specific difference as we expect it holds valuable information to probe the conditions close to the core of the star. We observe that our best model without overshooting ($X_0=0.7$ and $Z_0=0.017$), shown in orange, produces a small difference for the g-dominated mode, denoted $d_{02,\textrm{g}}$, which is about $\sim 1\mu\textrm{Hz}$ too large compared to the observed values (in blue). Now considering our best model with overshooting (in red), this difference significantly decreases by about $\sim 0.5\mu\textrm{Hz}$, placing it halfway between the model without overshooting and the observed differences. This strongly suggests that $d_{02,\textrm{g}}$ constitutes a relevant constraint on the overshooting parameter. Additionally, as the value of $d_{02,\textrm{g}}$ seems impacted by the convective mixing prescriptions, we computed models with the revised value of $\alpha_{\textrm{MLT}}=1.73$ for both cases (with and without overshooting -- in purple and green respectively). We note that $d_{02,\textrm{g}}$ seems to be anti-correlated with the mixing-length parameter, indicating that only an increase of its value could help; however this would contradict the decrease predicted by hydrodynamic simulations \citep{2015A&A...573A..89M}. As the overshooting parameter seems to be the most viable parameter, we try to reproduce the observed $d_{02,\textrm{g}}$ by generating a series of optimal models, according to the same procedure as previously (adjusting $\Delta\nu_0$, $\Delta\Pi_1$, and $\epsilon_{g,1}$), while varying the overshooting parameter between $0$ and $0.2$ with a step of $0.05$. This procedure is illustrated in Fig.~\ref{Fig:l2MM} where the black continuous line corresponds to the observed value of $d_{02,\textrm{g}}$, the dashed lines are the uncertainties, and each point is an optimal model with a given value of the overshooting parameter. We observe that the theoretical differences bracket the observed value for overshooting parameters of $0.15$ and $0.20$. By means of a linear interpolation, we found that a value of $\alpha_{\textrm{ov}}=0.17$ allows us to reproduce the observed difference. This point is also represented in Fig.~\ref{Fig:l2MM} where we observe a tight agreement between our model and the measured $d_{02,\textrm{g}}$.

\begin{figure}
    \centering
    \includegraphics[width=.95\linewidth]{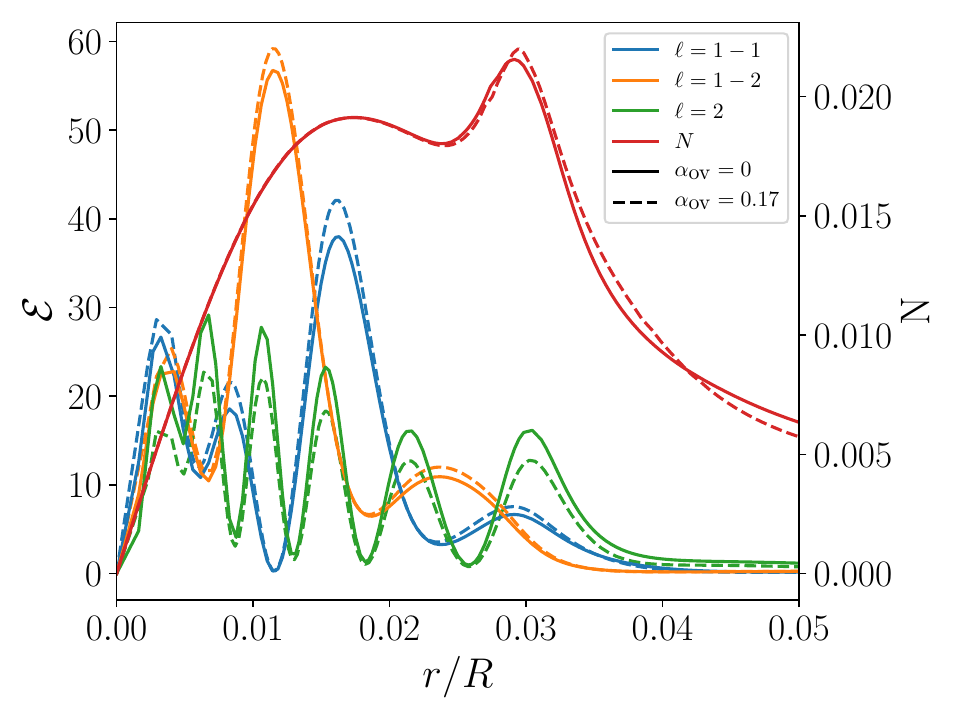}
    \caption{Energy density (Eq.~\ref{Eq:NRJ}) (normalised to $1$) as a function of $r$ of the two dipolar g-dominated mixed modes: $\nu \sim 750~\mu\textrm{Hz}$, denoted $\ell=1-1$ (in blue) and $\nu \sim 950~\mu\textrm{Hz}$, denoted $\ell=1-2$ (in orange), and of the only quadrupolar g-dominated mixed mode, denoted $\ell=2$ (in green). The Brunt-Väisälä frequency is also shown in red. These quantities are displayed for two of our best models, either without convective overshooting (continuous lines) or with the optimised overshooting parameter of $\alpha_{\textrm{ov}}=0.17$ (dashed line).}
    \label{Fig:NRJ}
\end{figure}

To better understand the probing potential of the quadrupolar g-dominated mixed mode, we plot the energy density of each identified g-dominated mode in Fig.~\ref{Fig:NRJ}. We define the energy density of a mode as:

\begin{equation}
    \mathcal{E} = \left[\xi_r^2+\ell\left(\ell+1\right)\xi_h^2\right] \rho r^2,\label{Eq:NRJ}
\end{equation}
with $\xi_r$ (resp. $\xi_h$) the radial (resp. horizontal) displacement of the mode (up to a multiplicative constant), $\rho$ the local density of the star, and $r$ the distance from the stellar center. The energy density is normalised so that the integral over $x=r/R$ is equal to $1$. 

In Fig.~\ref{Fig:NRJ}, we give the energy density of the first dipolar g-dominated mixed mode ($\nu \sim 750~\mu\textrm{Hz}$, denoted $\ell=1-1$, in blue), of the second dipolar g-dominated mixed mode ($\nu \sim 950~\mu\textrm{Hz}$, denoted $\ell=1-2$, in orange), and that of the only quadrupolar g-dominated mixed mode ($\nu \sim 750~\mu\textrm{Hz}$, denoted $\ell=2$, in green). We also show the Brunt-Väisälä frequency in the same regions. These quantities are shown for our best models without overshooting (continuous curves) and with the optimised overshooting parameter (dashed curves). We first observe that these three modes indeed have changing amplitudes in different parts of the stellar core, therefore probing these regions with various efficiencies. In particular, we note a correspondence between the peak left in $N$ by the H-shell and the energy density of the quadrupolar g-dominated mixed mode. Furthermore, the width of the energy density peak of this mode in the H-shell is comparable to the bump in $N$ at the same position. Consequently, this mode is particularly well suited to probe the H-shell. Furthermore, as we have noted in Sect.~\ref{Sec:Ov} (Fig.~\ref{Fig:mu}), the H-shell leaves a signature in the composition gradient that translates into the $N$ profile and varies with the overshooting parameter. As a consequence, the combined information carried by the dipolar and quadrupolar mixed modes constitute an opportunity to constrain the mixing history at the boundary of the core. 

\subsection{Impact of the surface effects}\label{Sec:Surf}
\begin{figure}
    \centering
    \includegraphics[width=.95\linewidth]{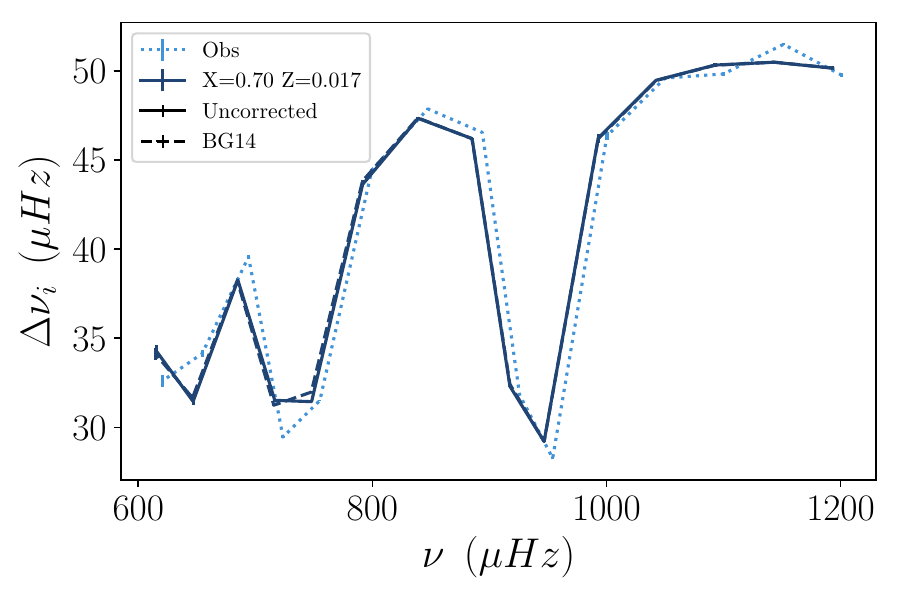}
    \caption{Pair-wise frequency separations for models with either uncorrected frequencies (black continuous line) or corrected frequencies according to \citet{2014A&A...568A.123B} (black dotted line) compared with the observed separations (blue dotted line).}
    \label{Fig:Surf}
\end{figure}

\begin{figure}
    \centering
    \includegraphics[width=.95\linewidth]{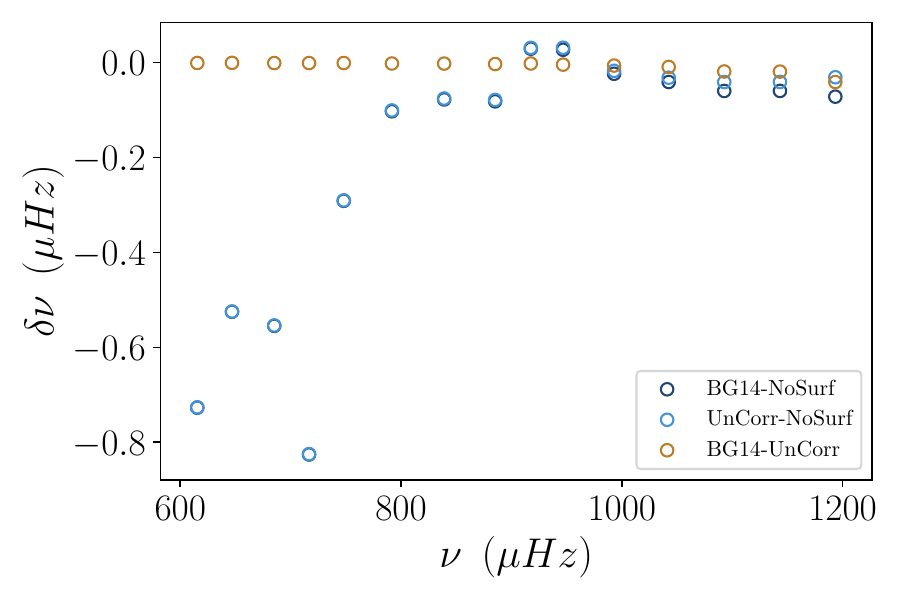}
    \caption{Frequency differences as a function of frequency. The difference between the frequencies of the optimal model including a correction of the surface effects following \citet{2014A&A...568A.123B} (BG14 in Table.~\ref{Tab:AllRes}) and that of the model without a correction for the surface effects ($X_0=0.70$, $Z_0=0.017$, and $\alpha_{\textrm{ov}}=0$ in Table.~\ref{Tab:AllRes}) is shown in dark blue. The difference between the same two models (BG14 and $X_0=0.70$, $Z_0=0.017$, and $\alpha_{\textrm{ov}}=0$) but using the frequencies of the corrected optimal model prior correction is shown in light blue. The light brown markers correspond to the magnitude of the surface correction term.}
    \label{Fig:SurfDiff}
\end{figure}

Finally, it has been established that high-frequency modes, which are the most affected by the stellar surface layers, are themost difficult to model due to the poor treatment in current one-dimensional stellar models of convection and its coupling with stellar oscillations. This coupling results from the comparable oscillation timescale with respect to the convective turnover time, leading to the adiabatic approximation to fail. However, these high frequency modes also are the ones with the largest measured uncertainties and these surface effects could have a negligible impact on our modelling results. To test this hypothesis, we computed one last model, starting from the composition ($X_0=0.70$ - $Z_0=0.017$) without overshooting, for which the theoretical frequencies have been corrected according to the `inverse-cubic' prescription from \cite{2014A&A...568A.123B}. To determine the values of the two coefficients of the \cite{2014A&A...568A.123B} prescription, we relied on the adjustment introduced by \citet[expressed in Eq.~20 and listed \href{https://www.aanda.org/articles/aa/pdf/2018/12/aa33783-18.pdf\#table.2}{Table~2}]{2018A&A...620A.107M} which relates them to model quantities, namely, the large frequency separation, effective temperature, surface gravity, and opacity. The individual frequency differences of the optimal model corrected for the surface, compared with our reference and the observations are shown in Fig.~\ref{Fig:Surf}. We note no significant change in the spectrum from the reference. The optimal parameters of this model are provided in Table~\ref{Tab:AllRes}, with the \emph{BG14} label in the \emph{Surf.} column. We note that, compared to its uncorrected counterpart there is only a change of $0.03\%$ in mass and $0.1\%$ in age. Therefore, in Gemma's case, the surface effects can safely be disregarded. Finally, we show in Fig.~\ref{Fig:SurfDiff} the differences between the frequencies of the optimal model including a correction of the surface effects and that of the model without a correction for the surface effects (second and last rows of Table \ref{Tab:AllRes}, in dark blue). We also show the same difference but using the frequencies of the corrected optimal model prior to BG14 correction (light blue). Finally, we also show the magnitude of the computed surface term in light brown. We observe that the two models are extremely similar and that the surface effects correction only slightly impacts the high-frequency modes. However, most of the relevant information used in this study is carried by low-frequency modes. This explains the negligible impact on the optimal model retrieved.

\section{Discussion}\label{Sec:Dis}
Using the information carried by Gemma's mixed-mode pattern, processed through EGGMiMoSA, we were able to build a large number of precise models (16). Among these models, there are two main families of modes (with or without overshooting). This is particularly interesting as Gemma sits close to the expected limit that separates stars with a radiative core ($M\lesssim1.2~M_{\odot}$) from those with a convective core ($M\gtrsim1.2~M_{\odot}$) on the MS. Therefore, it makes for a perfect candidate in the search for seismic tracers discriminating among these two scenarios, which greatly impact the stellar evolution up to the subgiant phase. Comparing models with the same initial composition but different overshooting parameters, we observe that when including overshooting with $\alpha_{\textrm{ov}}=0.15$, there is a systematic increase in mass by about $0.04~M_{\odot}$  (i.e. $3\%$ relative variation) and decrease in age by about $0.73~\textrm{Gyr}$ ($12\%$). This change is comparable to the requirements in precision of the future mission PLATO \citep{2014ExA....38..249R}: $15\%$ in mass and $10\%$ in age. The main responsible for the observed variation is the presence (respectively, absence) of a convective core during the MS in models with (resp. without) significant overshooting. This core leaves a sharper composition gradient at the H-shell that, in turn, reinforces the bump in the Brunt-Väisälä frequency which can be efficiently probed by the only observed quadrupolar g-dominated mixed mode. The structures of both models (with and without overshooting) right at the end of the MS are displayed in Fig.~\ref{Fig:nucTAMS}. From this figure, it is clear that the composition gradients (in light blue) are vastly different. Indeed, because models with overshooting (dashed curves) retain a convective core on the MS, they present a steep composition gradient left with the contraction of the core. While models without overshooting are not able to preserve a convective core and their composition gradient is less steep. These difference directly translate in the Brunt-Väisälä frequency (Eq.~\ref{Eq:Brunt}). Furthermore, by modelling another subgiant (KIC10273246), \citet{2021A&A...647A.187N} had noted that the presence of two dipolar g-dominated mixed modes already sufficed to put constraints on the central density and size of the main-sequence convective core. This is essential as it allows us to constrain the overshooting parameter and improve the accuracy of the retrieved stellar parameters. We obtained a value of $\alpha_{\textrm{ov}}=0.17$, which is rather high when compared to the only literature value (to our knowledge) of $0.074^{+0.017}_{-0.012}$ \citep{2024ApJ...965..171L}. Our value for the overshooting parameter is actually greater than any of the values they retrieved over the sample of stars they studied. If we now compare with the results of \cite{2023A&A...676A..70N} for two subgiants, we observe that our value lies in between theirs. Put in perspective with other studies, our value is definitely higher than expectations. For example, modelling double-lined eclipsing binaries, \citet{2016A&A...592A..15C,2017ApJ...849...18C} found a trend relating the stellar mass and the amount of overshooting over the HR diagram. They observe a sharp increase from $\alpha_{\textrm{ov}}=0.0$ to $\alpha_{\textrm{ov}}=0.2$ between $1.2 M_{\odot}$ and $2.0 M_{\odot}$. This is also incompatible with our results which, if we were to infer Gemma's mass according to the obtained overshooting parameter of $\alpha_{\textrm{ov}}=0.17$ and the observed trend noted by \citet{2016A&A...592A..15C,2017ApJ...849...18C}, would place it at a mass of about $1.8 M_{\odot}$. This result is obviously significantly larger than the value we obtained. However, we note that most of the stars in their sample belong to the small and large Magellanic cloud. The authors pointed out that these are notably less metallic than stars in the solar neighborhood. This decrease in metallicity might be a clue interpreting the fact that we obtained an overshooting parameter that does not agree with their observed trend. From our computations alone, a reduction of the metals mass fraction leads to a decrease in optimal mass, which would tend to favour reduced overshooting. However, additional computations would be necessary to explore this hypothesis. We also add that, as demonstrated in Sect.~\ref{Sec:Ov}, the main physical quantity that was probed is the composition gradient at the H-shell. While we found that a non-negligible overshooting value could serve as a mechanism to produce the expected composition gradient, this is not the only possible explanation. For example, modified high temperature opacities or nuclear reaction cross-sections could also affect the chemical composition profiles and possibly produce similar results. Nevertheless, testing these types of processes is beyond the scope of the present paper.

When comparing with other authors modelling Gemma, we tend to be on the more massive side. Indeed, our masses range between $1.146~M_{\odot}$ and $1.257~M_{\odot}$ with the two more extreme compositions. On the other hand, \citet{2024ApJ...965..171L} only retrieved a mass of $1.09^{+0.06}_{-0.05}~M_{\odot}$ and \citet{2020MNRAS.495.2363L} obtain $1.18\pm 0.08~M_{\odot}$. The source of this discrepancy remains unclear, but it is of a comparable magnitude as that produced by a change in the stellar evolution code can produce \citep[see e.g.][\href{https://www.aanda.org/articles/aa/pdf/2014/04/aa22779-13.pdf\#table.12}{Table 12}]{2014A&A...564A..27D}. From our results alone (see Table~\ref{Tab:AllRes}), we may infer two pieces of information. First, at constant $X_0$, an increase in $Z_0$ translates to an increase in optimal mass. Second, at constant $Z_0$, an increase in $X_0$ also leads to an increase in optimal mass, as well as in the optimal age. This provides a plausible explanation for the smaller mass obtained by \citet{2024ApJ...965..171L}. Indeed, their study incorporated the abundances determined by \citet{1998SSRv...85..161G} ($Z/X=0.0231$). We used the more recent and less metallic one provided by \citet{2009ARA&A..47..481A} ($Z/X=0.0181$), their initial metallicity is significantly lower than ours. They determined $\left[Fe/H\right]_0=-0.13\pm 0.11$, which translates to $\left(Z/X\right)_0=0.017$, using the value of $Z/X=0.0231$ quoted earlier. Paired with their optimal initial helium abundance of $Y_0=0.279\pm0.02$, we found $X_0=0.709$ and $Z_0=0.012$. This is lower than the lower bound of range of values we tested, namely, $Z_0\in\left[0.013,0.017\right]$, and we would expect that a reduction of the metals mass fraction leads to a reduction of the optimal mass. Nevertheless, our study favours a larger metals abundance as only a model with $X_0=0.70$ and $Z_0=0.017$ is able to account for all the constraints (seismic and non-seismic) considered in the present study. Our results provide a better agreement with those of \citet{2020MNRAS.495.2363L} who found $M = 1.18 \pm 0.08~M_{\odot}$ and $t=5.69 \pm 0.68~\textrm{Gyr}$. This statement also holds true when comparing the luminosities and radii we retrieve with both studies (see Table~\ref{Tab:AllRes}). We observe that our values lie on the upper side of the range determined by \citet{2020MNRAS.495.2363L}, $R=2.06 \pm 0.04 R_{\odot}$ and $L=3.80 \pm 0.25 L_{\odot}$, while \citet{2024ApJ...965..171L} derived values on the lower end of this range, $R=2.01^{+0.04}_{-0.03} R_{\odot}$ and $L=3.42^{+0.25}_{-0.23} L_{\odot}$. This illustrates biases in our respective studies. The origin of which is most likely related to differences in the elected methodologies but remains uncertain. Additionally, it was noted by \citet{2021A&A...647A.187N} that the knowledge of $\Delta\nu$ and a single dipolar g-dominated mixed mode was sufficient to constrain the central density, $\rho_c$, down to $1\%$ precision. Our results are in relative agreement as we observe that the maximum spread in central density retrieved in our models is of only $2\%$ (see Table~ \ref{Tab:AllRes}). We also note that the ratio of the central density over the mean density of the star, $\rho_c/\overline{\rho}$, is even better constrained, with a spread of only $0.8\%$. This clearly suggests that the selected indicators carry strong structural constraints.

In Table~\ref{Tab:AllRes}, we do not quote any of the uncertainties on the individual models. This stems from the optimisation procedure we selected, which we find does not provide reliable uncertainties. Indeed, these are estimated via the inverse of the Hessian matrix of the problem. As a consequence, they correspond to the error propagation of the observed uncertainties on the seismic indicators (which are propagated from the observed frequencies). Consequently, they are extremely small, yielding $\sigma\left(t\right)\sim 0.2\cdot 10^{-3}~\textrm{Gyr}$ ($\sim 10^{-14}$ relative error) and $\sigma\left(M\right)\sim 0.1\cdot 10^{-4}~M_\odot$ ($\sim 10^{-5}$ relative error), thus unrealistic. This is further reinforced by the fact that we used only two free parameters. As such, the computed uncertainties do not account for the latitude provided by additional degrees of freedom. Nevertheless, we tested several possibilities for the chemical composition and overshooting parameter. By averaging over all the results from Table~\ref{Tab:AllRes} and computing the spread between the average values and the extrema, it is possible to provide a more conservative determination of the uncertainties: $M=1.19^{+0.07}_{-0.04}$, $t=5.75^{+0.54}_{-0.48}$.  While these uncertainties are more realistic, a better approach would be to include other free parameters in the fitting procedure (which tend to destabilise the model search) or make use of global search techniques, such as AIMS \citep{2019MNRAS.484..771R}, which are able to efficiently sample the parameter space, offering robust uncertainties. However, such techniques are beyond the scope of the present paper, since implementing our EGGMiMoSA technique in other fitting algorithms would take a substantial amount of time. Finally, we have not provided uncertainties on the optimised overshooting parameter as this parameter is not part of the free parameters of the optimisation procedure. From the variation in the $d_{02,\textrm{g}}$ difference over the several overshooting parameter values considered in Fig.~\ref{Fig:l2MM} it is possible to provide a rough estimation of this uncertainty. However, this yields too small and unrealistic uncertainties. Additionally, from the pool of models in the present paper, we may also observe a strong dependence of this difference on the composition. We expect that this dependence would further increase the size of the estimated uncertainties.

\section{Conclusions}
Taking advantage of the suite of tools we have developed for this purpose \citep[WhoSGlAd, EGGMiMoSA, and PORTE-CLES][Farnir \& Dupret 2025 in prep.]{2019A&A...622A..98F,2021A&A...653A.126F}, we have carried out the one of the first detailed seismic modelling of a subgiant star, Gemma (KIC11026764). We find that the large frequency separation of radial modes, the period spacing, and gravity offset of dipolar mixed modes, along with the amplitude of the helium glitch signature and the small separation between one quadrupolar g-dominated mixed mode and the closest radial mode provide strong structural constraints. Indeed, we observe that the ratio of central density to mean density of the star is very well constrained, down to $0.8\%$, not unlike the observations of \citet{2021A&A...647A.187N}. Additionally, thanks to the inclusion of the evaluation of the helium glitch amplitude (along with spectroscopic constraints) allows us to favour an initial composition of $X_0=0.70$ and $Z_0=0.017$. This is more metallic than what was obtained by \citet{2024ApJ...965..171L}, $\left(Z/X\right)_0=0.024$ compared to $0.017$. This results in more massive optimised models when compared to their work. 

We also noted that the presence of a quadrupolar g-dominated mixed mode provides strong constraints on the composition gradient at the edge of the H-shell. To reproduce the small separation associated with this mixed mode, we need to fine-tune the central overshooting in the MS. We retrieved $\alpha_{ov}=0.17$. While our value is relatively large compared to what we would expect from other studies \citep[see e.g. ][]{2016A&A...592A..15C}, the ability to measure this parameter will be crucial to improve our understanding of the mixing history of solar-like stars. However, we must insist that the inclusion overshooting is only one of the possible process to account for the composition gradient, which is the key quantity probed in this context. The present study will serve as a pilot study for a broader work exploring a larger sample of subgiant stars with a broad variety of masses and at different evolutionary stages. The Kepler Legacy sample \citep{2017ApJ...835..172L} and the future PLATO mission \citep{2024arXiv240605447R} will provide ample data for such studies.

\begin{acknowledgements}
The authors would like to thank the referee for their constructive remarks that helped improve the present paper clarity and readability.\\
M.F. is a Postdoctoral Researcher of the Fonds de la Recherche Scientifique – FNRS.\\
GB acknowledges fundings from the Fonds National de la Recherche Scientifique (FNRS) as a postdoctoral researcher.
\end{acknowledgements}

\bibliographystyle{aa}
\bibliography{bibliography}

\appendix

\end{document}